\def\imo{i}

\def\imo{i}
\def\rem#1{}

\def\be{\begin{equation}}
\def\ee{\end{equation}}
\def\bea{\begin{eqnarray}}
\def\eea{\end{eqnarray}}

\def\Re#1{\mathrm{Re}(#1)}

\def\imo{i}

\documentclass[11pt]{article}
\usepackage[dvips]{graphicx}
\usepackage{amsmath,amssymb,bm}
\begin{document}

\title{\textbf{
	Evolution of perturbations of squashed Kaluza-Klein black holes:
	escape from instability\\
}
\vspace{1cm}
}

\author{Hideki Ishihara, Masashi Kimura,\\
\emph{Department of Mathematics and Physics, Graduate School of Science} \\
\emph{Osaka City University, Osaka 558-8585, Japan}\\  \\
Roman A. Konoplya, Keiju Murata,\\
\emph{Department of Physics, Kyoto University, Kyoto 606-8501, Japan} \\ \\
Jiro Soda, \\
\emph{Department of Physics, Kyoto University, Kyoto 606-8501, Japan}, \\
\emph{Kavli Institute for Theoretical Physics}, \\
\emph{Zhong Guan Cun East Street 55, Beijing 100080, P.R. China}\\ \\
and\\ \\
Alexander Zhidenko \\
\emph{Instituto de F\'{\i}sica, Universidade de S\~{a}o Paulo}, \\
\emph{C.P. 66318, 05315-970, S\~{a}o Paulo-SP, Brazil}}

\date{}

\maketitle
\vspace{2cm}

\thispagestyle{empty}

PREPRINT:~{OCU-PHYS 290},~{AP-GR 55},~KUNS-2122,~CAS-KITPC/ITP-020

\begin{abstract}
The squashed Kaluza-Klien (KK) black holes differ from the Schwarzschild black holes
with asymptotic flatness or the black strings even at energies
for which the KK modes are not excited yet, so that squashed KK black holes open
a window in higher dimensions.
Another important feature is that the squashed KK black holes
are apparently stable and, thereby, let us avoid the Gregory-Laflamme instability.
In the present paper, the evolution of scalar and gravitational perturbations
in time and frequency domains is considered for these squashed KK black holes.
The scalar field perturbations are analyzed for general rotating squashed KK
black holes. Gravitational perturbations for the so called zero mode are shown
to be decayed for non-rotating black holes, in concordance with
the stability of the squashed KK black holes.
The correlation of quasinormal frequencies with the size of extra dimension
is discussed.
\end{abstract}

\maketitle

\newpage
\section{Introduction}

Recent years, higher dimensional black holes have become one of the key objects of
the modern high energy physics.
In brane world scenarios with large extra dimensions\cite{ADD}, it was
suggested that mini black holes may be created in particle
colliders\cite{accelerator}, which  encouraged thereby
detailed investigation of behavior of
particles and fields around the black holes.

Among important processes, the proper oscillations
of higher dimensional black holes as a response to an external perturbations
have been actively studied recently \cite{DBH_QNMs1}-\cite{DBH_QNMs15}.
These oscillations are governed by the so-called quasinormal modes,
which have been in the focus of gravitational
research during recent years because of a few reasons.
First, quasinormal modes are expected to be an observed fingerprint of
a black hole's gravitational waves with the help of a new generation of
gravitational antennas. Second, for some
asymptotically anti de Sitter space-times the quasinormal modes have
interpretation in the dual conformal field theory (CFT) \cite{QNM_AdS_CFT},
and help thereby to study some finite temperature field theory processes, such as hydrodynamics of quark-gluon plasma \cite{Son:2007vk} in CFT.
Finally, complete numerical investigation of quasinormal modes makes it possible to prove (in)stability for some black holes,
when analytical proof is difficult \cite{Ishihbashi-Kodama, 1, Murata:2007gv}.

Usually, the asymptotic flatness of black hole spacetimes,
which is an idealization of the boundary condition for isolated systems,
is required in four dimensions.
In higher dimensions, however, since the extra dimensions are expected to be
compactified into a small scale, so-called Kaluza-Klein geometry,
then the higher dimensional black holes have the asymptotic structure
of the Kaluza-Klein type
\footnote{  
Asymptotically flatness would be a reasonable assumption for higher dimensional
black holes if the size of horizon is sufficiently smaller than
the size of extra dimensions.}.
It is interesting to study how the compactness of the extra dimension affects
the quasinormal modes of the Kaluza-Klein black holes.

The simplest example of five-dimensional black objects with
the Kaluza-Klein geometry is the black string,
 the direct product of four-dimensional black hole and a circle.
These objects look different from four-dimensional black hole
only at sufficiently high energies, when Kaluza-Klein modes are
excited \cite{Gregory:1993vy, Kanti:2004nr,rev,MSK}.
Therefore within these space-times we need high
energy regime to see the extra dimension.

If we allow the asymptotic structure of twisted S$^1$ bundle over
four-dimensional Minkowski space-time,
there exist exact solutions of Kaluza-Klein black holes
with squashed horizons for
 neutral~\cite{DM, Wang}, charged~\cite{Ishihara:2005dp, NIMT},
and extremely charged cases~\cite{superBH, IKMT1}.
%
%
Such black holes look like five-dimensional squashed black holes
near the event horizon, and like a Kaluza-Klein space-time at spatial infinity,
i.e., locally a direct product of four-dimensional Minkowski
space-time and a circle. Physical properties of the squashed Kaluza-Klein
black holes are studied in \cite{ThermoDynamics, DBH_kucha3,CWS},
and generalizations of them appear in \cite{Generalizations}.
Owing to the non-trivial bundle structure, the size of the extra dimension might
be observed even at low energies\cite{DBH_kucha3}.
Then, it is interesting to study the the quasinormal modes of
the squashed Kaluza-Klein black holes.
This is even more motivated if we take into account that squashed Kaluza-Klein
black holes are gravitationally stable \cite{Soda1}.

In the present paper we study evolution of perturbations of squashed Kaluza-Klein
black holes
both in time and frequency domains and find the quasinormal spectrum for scalar field and gravitational perturbations
of these black holes.

 The paper is organized as follows: in Sec II we study the quasinormal modes of scalar field perturbations
around rotating Kaluza-Klein black holes with squashed horizon. Sec III is devoted to gravitational perturbations of non-rotating
black holes. Finally in Sec. IV we discuss the obtained results.

\section{Quasinormal modes of the scalar field for rotating squashed
Kaluza-Klein  black holes}

In this section, we investigate the quasinormal modes of the scalar field
in the rotating uncharged squashed Kaluza-Klein black holes.

The five-dimensional rotating squashed KK black hole with two equal angular momenta is described by\cite{DM, Wang}
\begin{eqnarray}
ds^2=-dt^2+\frac{\Sigma_0}{\Delta_0}k(r)^2dr^2+\frac{r^2+a^2}{4}
[k(r)(\sigma^2_1+\sigma^2_2)+\sigma^2_3]+\frac{\mu}{r^2+a^2}(dt-\frac{a}{2}\sigma_3)^2,
\label{metric}
\end{eqnarray}
with
\begin{eqnarray}
\sigma_1 = -\sin{\psi} d\theta+\cos{\psi} \sin{\theta}d\phi \ , \quad
\sigma_2 = \cos{\psi} d\theta+\sin{\psi} \sin{\theta}d\phi \ , \quad
\sigma_3 = d\psi+\cos{\theta}d\phi \ ,
\end{eqnarray}
where $0<\theta<\pi$, $0<\phi<2\pi$ and $0<\psi<4\pi$. The parameters are given by
\begin{eqnarray}
\Sigma_0&=&r^2(r^2+a^2),\nonumber\\
\Delta_0&=&(r^2+a^2)^2-\mu r^2,\nonumber\\
k(r)&=&\frac{(r^2_{\infty}-r^2_+)(r^2_{\infty}-r^2_-)}{(r^2_{\infty}-r^2)^2}.
\end{eqnarray}
Here $\mu$ and $a$ are parameters which correspond to mass and angular momenta,
respectively. $r=r_+$ and $r=r_-$ are outer and inner horizons of the black hole
and they relate to $\mu$ and $a$ by $a^4=(r_+r_-)^2, \mu-2a^2=r^2_++r^2_-$.
The parameter $r_{\infty}$ corresponds to the spatial infinity.
In the parameter space $0 < r_-\leq r_+ < r_{\infty}$, $r$ is restricted within
the range $0<r<r_{\infty}$. The shape of black hole horizon is deformed by
$k(r_+)$.

The wave equation for the massless scalar field $\Phi(t,r,\theta,\phi,\psi)$ in the background (\ref{metric}) obeys
\begin{eqnarray}
\frac{1}{\sqrt{-g}}\partial_{\mu}(\sqrt{-g}g^{\mu\nu}\partial_{\nu})
\Phi(t,r,\theta,\phi,\psi)=0.\label{WE}
\end{eqnarray}
Taking the ansatz $\Phi(t,r,\theta,\phi,\psi)=e^{-i\omega t}R(\rho)e^{i m\phi+i\lambda \psi}S(\theta)$, where $S(\theta)$ is the so-called spheroidal harmonics, the radial and time variables can be decoupled from angular ones, so that the final wave-like equation reads

\begin{eqnarray}
\frac{d}{d\rho}\bigg[\Delta\frac{d R(\rho)}{d\rho}\bigg]
+\bigg[\frac{\tilde{H}^2}{\Delta}+\Lambda-l(l+1)+\lambda^2\bigg]R(\rho)=0,\label{radial}
\end{eqnarray}
where $l$ is the non-negative integer multipole number, $|m|<l$ and $|2\lambda|<2l$
are integers, and
\begin{eqnarray}
&&\tilde{H}^2=\frac{\mu r^2_{\infty}(\rho+\rho_0)^4}{H^4(r^2_{\infty}+a^2)^2}
\bigg[\omega-\frac{\lambda a
H^2(r^2_{\infty}+a^2)}{\rho_0r^3_{\infty}}\bigg]^2,
\\
&&\Lambda=\frac{4\rho^2_0r^6_{\infty}(\rho+\rho_0)^2}{H^2(r^2_{\infty}+a^2)^4}\omega^2-
\frac{4\lambda^2(\rho+\rho_0)^2}{r^2_{\infty}+a^2},
\\
&&H^2=\frac{\rho+\rho_0}{\rho+\frac{a^2}{r^2_{\infty}+a^2}\rho_0}.
\end{eqnarray}
The radial coordinate $\rho$ is given by
\begin{eqnarray}
\rho=\rho_0\frac{r^2}{r^2_{\infty}-r^2},\label{p}
\end{eqnarray}
with
\begin{eqnarray}
\rho^2_0&=&\frac{k_0}{4}(r^2_{\infty}+a^2),\nonumber \\
k_0&=&k(r=0)=\frac{(r^2_{\infty}+a^2)^2-\mu r^2_{\infty}}{r^4_{\infty}}.
\end{eqnarray}
Note that the three parameters $\rho_0$ and $\rho_\pm=\rho_0r_\pm^2/(r^2_{\infty}-r_\pm^2)$ can define the metric (\ref{metric}) if $r_\infty<\infty$.
In some papers they are used to parameterize the black hole instead of the parameters $r_\infty$, $r_\pm$.

In frequency domain we used the following Frobenius expansion
\begin{equation}\label{Frobenius}
R = \left(\frac{r^2-r_+^2}{r^2-r_-^2}\right)^{-\imo\kappa}e^{\imo\rho\Omega}\rho^{\imo\nu-1}\sum_{n=0}^\infty a_n\left(\frac{r^2-r_+^2}{r^2-r_-^2}\right)^n,
\end{equation}
where $\kappa$, $\nu$ and $\Omega$ are chosen in order to eliminate the singularities at $r=r_+$ and $r=r_\infty$. The sign of $\kappa$ and $\Omega$ is
chosen in order to remain them in the same complex quadrant as $\omega$.
We substitute (\ref{Frobenius}) into (\ref{radial}) and obtain the recurrence relation for the coefficients $a_n$.
After the recurrence relation is known we can find the equation with the infinite continued fraction with respect to
$\omega$, which can be solved numerically \cite{continuedfraction}.

\begin{figure}[htbp]
 \centering
 \includegraphics[scale=0.5]{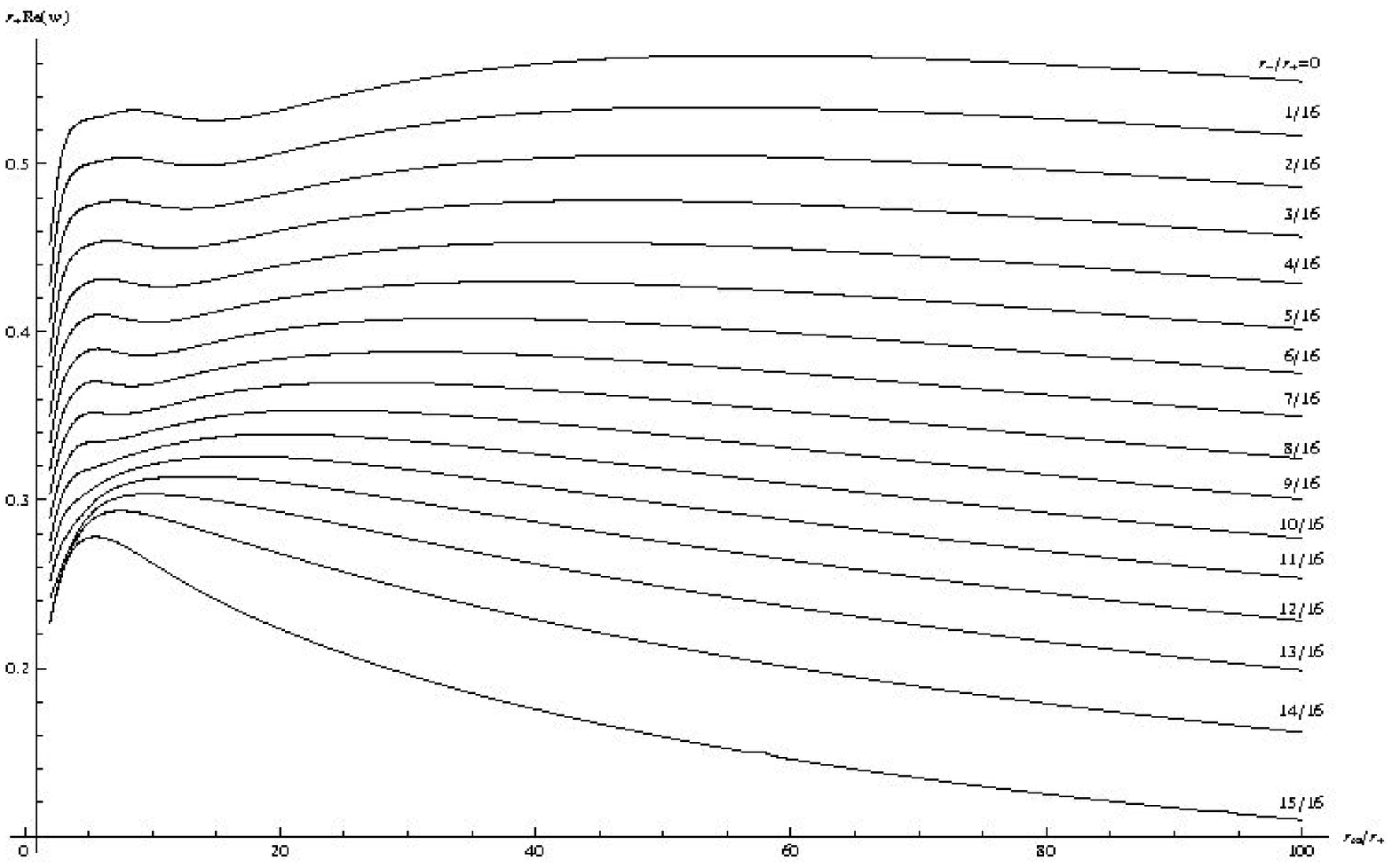}
 \caption{
%
Real parts of quasinormal frequencies as a function of $r_-$ and $r_\infty$ ($l=0$).
}
 \label{Real_part}
 \centering
 \includegraphics[scale=0.5]{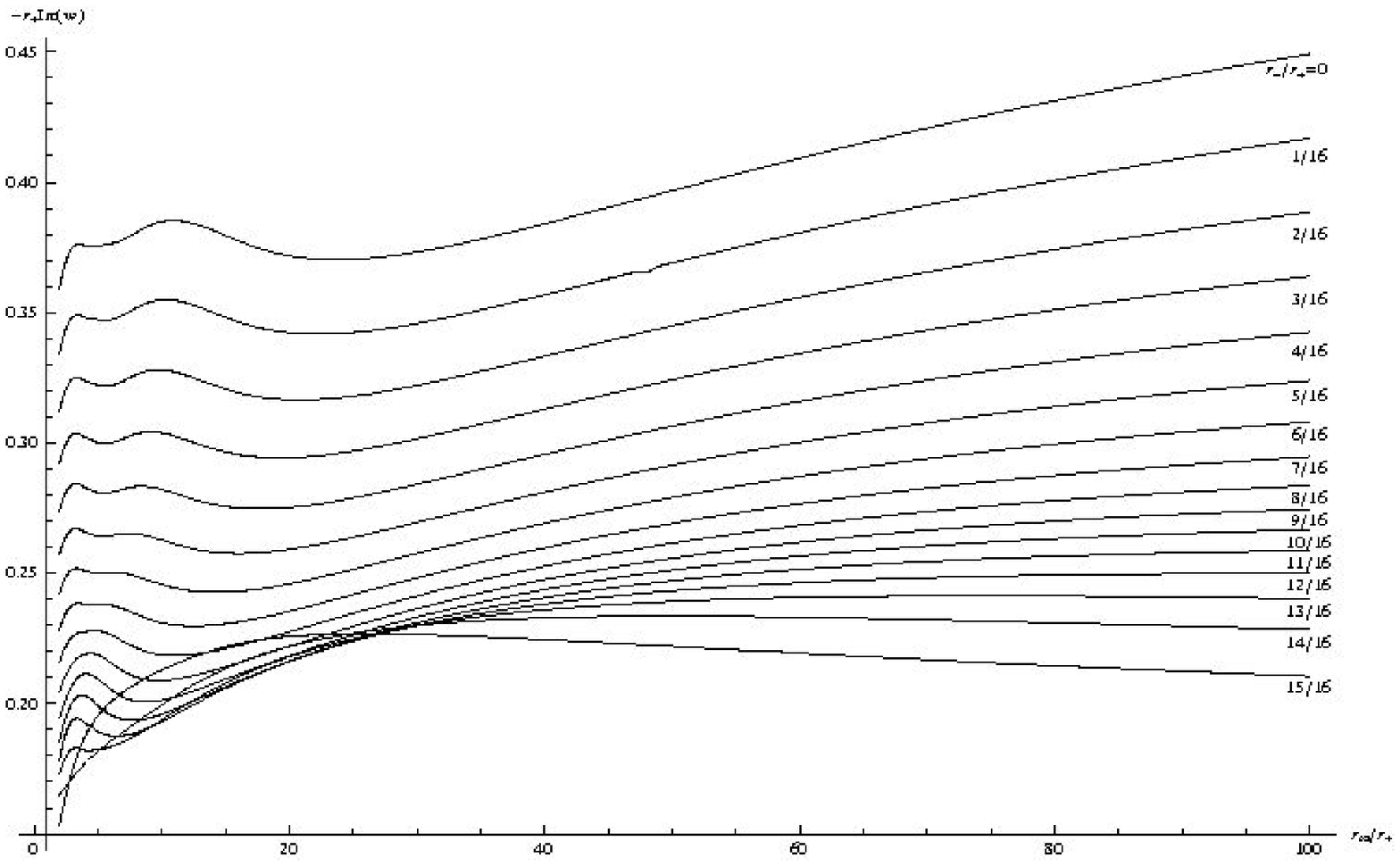}
 \caption{
Imaginary parts of quasinormal frequencies as a function of $r_-$ and $r_\infty$
($l=0$).}
 \label{Imaginary_part}
\end{figure}

The fundamental quasinormal modes will be of our primary interest hereafter, when dealing with frequency domain, because fundamental modes dominate
in a late time oscillations. Contribution of all overtones
can be seen in time domain considered in the next section. The fundamental modes for different values of $\lambda$ are shown in Fig.\ref{Real_part} and
Fig.\ref{Imaginary_part}, and tables 1 and 2.
There one can see that the real oscillation frequency exerts some irregular growth (with local minimums) when $r_{\infty}$ is increasing until some moderately
large values of $r_{\infty}$. At larger $r_{\infty}$ the growth of $\Re\omega$ changes into monotonic decay. The imaginary part of $\omega$ that determines
the damping rate also has some initial irregular growth when $r_{\infty}$ increases, but at larger $r_{\infty}$ the two scenarios are possible: either monotonic decay
(for large values of ratio $r_{-}/r_{+}$) or monotonic growth (for small and moderate $r_{-}/r_{+}$) (see Fig. 2).
Thus, for a given mass and angular momentum of the black hole,
one can learn the size of extra dimension $r_{\infty}$ from values of quasinormal modes of the emitted radiation.

The calculations performed with the help of the Frobenius method are accurate and WKB values given here are rather an additional check. From tables 1 and 2 we can
see that for small values of ratio $\rho_0/\rho_+$  the WKB method extended until
the 6th WKB order beyond the eikonal approximation \cite{Will-Schutz, WKB}
is in a  very good agreement with the Frobenius method,
but the larger $\rho_0/\rho_+$ the worse convergence of the WKB series.



%
%

\begin{table}
\begin{center}
\begin{tabular}{|c|c|c|c|}
\hline
$l$&$\lambda$&WKB (6th order)&Frobenius\\
\hline
$10$&$0$&  $3.5149-0.15956\imo$&$3.510564-0.159674\imo$\\
$10$&$1/2$&$3.5336-0.15821\imo$&$3.529225-0.158317\imo$\\
$10$&$1$&  $3.5898-0.15413\imo$&$3.585417-0.154231\imo$\\
$10$&$3/2$&$3.6842-0.14729\imo$&$3.679778-0.147368\imo$\\
$10$&$2$&  $3.8178-0.13759\imo$&$3.813421-0.137647\imo$\\
$10$&$5/2$&$3.9924-0.12493\imo$&$3.988012-0.124944\imo$\\
$10$&$3$&  $4.2103-0.10912\imo$&$4.205922-0.109076\imo$\\
$10$&$7/2$&$4.4747-0.08988\imo$&$4.470523-0.089745\imo$\\
\hline
\end{tabular}
\caption{Scalar field perturbations. Comparison of 6th-order WKB and
Frobenius approaches for the non-rotating uncharged squashed Kaluza-Klein
black holes ($\rho_0/\rho_+=3$, $r_\infty/r_+=2$).
Quasinormal frequencies are measured in units of $\rho_+$.}
\end{center}
\end{table}


\section{Gravitational quasinormal modes for non-rotating squashed
Kaluza-Klein  black holes}

In the previous section, we have considered the scalar field.
However, the tensor perturbations are more interesting from the point of
view of the stability and the gravitational waves from black holes.

The metric of the uncharged non-rotating squashed Kaluza-Klein black hole is a particular case of the previous metric
\begin{eqnarray}
ds^2 =-F(\rho)d\tau^2 + \frac{G(\rho)^2}{F(\rho)}d\rho^2
+ 4 \rho^2 G(\rho)^2  \sigma^+ \sigma^-
+\frac{r_{\infty}^2}{4 G(\rho)^2}(\sigma^3)^2,
\end{eqnarray}
where we have difined $\tau = 2\rho_0 t /r_\infty$ and
\begin{eqnarray}
F(\rho) = 1-\frac{\rho_+}{\rho} \ , \quad
G(\rho)^2 = 1+\frac{\rho_0}{\rho} \ , \quad
r_{\infty}^2 &=& 4 \rho_0(\rho_+ + \rho_0) \ .
\end{eqnarray}
Here, we have used a basis
\begin{equation}
 \sigma^{\pm} = \frac{1}{2}(\sigma^1 \mp i \sigma^2)\ .
\end{equation}

The perturbed metric is
\begin{eqnarray}
h &=& h_{AB}dx^Adx^B
        + 2h_{Ai}dx^A \sigma^i
        + h_{ij}\sigma^i \sigma^j
\end{eqnarray}
where indices run $A = t,\rho,~~i,j = +,-,3$.
Since the spacetime has the symmetry $SU(2)\times U(1)$, the metric perturbations
can be classified by eigenvalues $J,M$ for $SU(2)$ and $K$ for $U(1)$.
Here we consider only zero modes $J=M=0$. Even in this case,
since $\sigma^\pm $ carry eigenvalues $K=\pm 1$, each components could have different
eigenvalue $K$. It is important to recognize that the components with different
$K$ are decoupled. That is why we have the master equation for each $K$.
To obtain master equations, we choose the gauge condition as
\begin{eqnarray}
h_{3+} = h_{3-} = h_{+-} = h_{tt} = h_{t3}  =0
\end{eqnarray}
%
As is shown in \cite{Soda1}, the perturbation equations  for the $|K|=2$ mode
can be reduced to the wave equation for $h_{++}$
with the effective potential in the form
\begin{eqnarray}
V_2 &=& \frac{
         (\rho-\rho_+)}{16\rho^3 \rho_0 (\rho_++\rho_0)(\rho+\rho_0)^3}
\Big[64\rho^5 + 256\rho^4\rho_0-32\rho^3 \rho_0(\rho_+-11\rho_0)
\nonumber\\
  &&  \qquad           +8\rho^2\rho_0(2\rho_+^2 -5\rho_+\rho_0+25\rho_0^2)
                            +\rho \rho_0^2 (20\rho_+^2-9\rho_+\rho_0+35\rho_0^2)
+9\rho_+ \rho_0^3(\rho_+ +\rho_0)
                       \Big] \ .
\end{eqnarray}
%
%
Similary, the perturbation equations  for the $|K|=1$ mode
can be reduced to the wave equation for $h_{\rho +}$
with the effective potential
%
%
\begin{eqnarray}
V_1&=&\frac{\left( \rho  - \rho_+   \right)}
  {16 \rho ^3 \rho_0   {\left( \rho  + \rho_0   \right) }^3
    \left( \rho_+   + \rho_0   \right)
    {\left( \rho ^2 + 2 \rho  \rho_0   - \rho_+   \rho_0   \right) }^2}
\Big[
     16 \rho ^9 + 128 \rho ^8 \rho_0   -
      27 \rho_+  ^3 \rho_0  ^5 \left( \rho_+   + \rho_0   \right)
\nonumber\\
  &&  \qquad +
      32 \rho ^7 \rho_0   \left( \rho_+   + 15 \rho_0   \right)  +
      2 \rho ^2 \rho_+   \rho_0  ^4
       \left( 111 \rho_+  ^2 - 53 \rho_+   \rho_0   - 228 \rho_0  ^2
         \right)
\nonumber\\
  &&  \qquad + 8 \rho ^6 \rho_0
       \left( -16 \rho_+  ^2 - 9 \rho_+   \rho_0   + 119 \rho_0  ^2 \right)
          + \rho  \rho_+  ^2 \rho_0  ^4
       \left( -20 \rho_+  ^2 + 151 \rho_+   \rho_0   + 187 \rho_0  ^2
         \right)
\nonumber\\
  &&  \qquad + \rho ^4 \rho_0  ^3
       \left( -827 \rho_+  ^2 - 959 \rho_+   \rho_0   + 764 \rho_0  ^2
         \right)  + \rho ^5 \rho_0  ^2
       \left( -484 \rho_+  ^2 - 501 \rho_+   \rho_0   +
         1103 \rho_0  ^2 \right)
\nonumber\\
  &&  \qquad+
      2 \rho ^3 \rho_0  ^3 \left( 36 \rho_+  ^3 - 361 \rho_+  ^2 \rho_0   -
         495 \rho_+   \rho_0  ^2 + 126 \rho_0  ^3 \right)
\Big].
\end{eqnarray}
And, for the $K = 0$ mode, we obtain the wave equation for $h_{33}$
with the effective potential
\begin{eqnarray}
V_0 &=& \frac{(\rho-\rho_+)}{16\rho^3 (\rho+\rho_0)^3(4\rho+3\rho_0)^2}
\Big[256\rho_+\rho^4 + 64\rho^3 (17\rho_++2\rho)\rho_0 +48\rho^2(32\rho_+ +11\rho)\rho_0^2
\notag\\ &&  \qquad
+60\rho(13\rho_++12\rho)\rho_0^3+9(9\rho_+ +35\rho)\rho_0^4
\Big] \ .
\end{eqnarray}


We shall use the time-domain analysis based on the standard integration scheme for the wave-like equations
\begin{equation}\label{hyperbolic}
\left(\frac{\partial^2}{\partial t^2}-\frac{\partial^2}{\partial \rho_\star^2}\right)\Psi_i (t,\rho)
=-V_i (\rho)\Psi_i (t,\rho) \ , \quad (i=1,2,3)  \ ,
\end{equation}
which is described for instance in \cite{Price-Pullin}.
In detail, we applied a numerical characteristic integration scheme,that uses the light-cone variables $u = t - \rho_\star$ and $v
= t + \rho_\star$. In the characteristic initial value problem, initial data are specified on the two null surfaces $u = u_{0}$ and $v = v_{0}$. The discretization scheme we used, is
\begin{equation}\label{d-uv-eq}
\Psi(N) = \Psi(W) + \Psi(E) - \Psi(S) -\Delta^2\frac{V(W)\Psi(W) + V(E)\Psi(E)}{8} + \mathcal{O}(\Delta^4) \ ,
\end{equation}
where we have used the following definitions for the points: $N =(u + \Delta, v + \Delta)$, $W = (u + \Delta, v)$, $E = (u, v + \Delta)$ and $S = (u,v)$.
We also used in this paper the WKB method developed in \cite{Will-Schutz} and extended to the sixth order in \cite{WKB}. As it is described in too many papers recently
we refer a reader to the above mentioned papers for a detailed description.

The quasinormal frequencies were obtained by WKB formula (Fig. 3) and by the above time domain integration (Table 3.). There are two distinctive features
of gravitational perturbations, on the contrary to test scalar field perturbations.
First, the $K=1$ potential contains the small negative gap near the event horizon,
which nevertheless does not induce any instability \cite{Soda1}. Second, $K=1, 2$
perturbations for some values of $\rho_{0}$ have effective potentials without
maximum but with a monotonic growth instead. For the latter case the WKB method
cannot be applied but the time domain integration can be.

From Fig. 4, 5 and 6, one can see that the quasinormal ringing for $K=0$ case
and for $K=1,2$ cases are completely different.
For $K=0$ perturbations one can see the usual damped oscillations until sufficiently
large time, when they are dominated by the asymptotic
late-time  tails (see Fig. 4).
For $K=1, 2$ cases we see that "tail behavior" occurs at much earlier time
(see Fig. 5 and 6),
as it takes place for perturbations of massive scalar field
\cite{Masive_Scalar}. It is not surprise if one remembers that the effective potential for the massive scalar field with a large mass also has monotonic behavior as
in our cases for $K=1, 2$ gravitational perturbations.

\begin{table}
\begin{center}
\begin{tabular}{|c|c|}
\hline
$\rho_{0}$ & $\omega$ \\
\hline
0 & $\omega = 0.226-0.1711 \imo $ \\
0.1 & $\omega = 0.234-0.2046\imo$ \\
0.2 & $\omega = 0.243-0.2003$ \\
0.5 & $\omega = 0.263-0.1848\imo$  \\
1 & $\omega = 0.274-0.1635\imo$ \\
2 & $\omega = 0.271-0.1344\imo$ \\
\hline
\end{tabular}
\caption[smallcaption]{Perturbations of the $K=0$ type. Quasinomal frequencies
for small $\rho_{0}$.
}
\label{label}
\end{center}
\end{table}


For larger values of $\rho_{0}$ both real and imaginary parts of $\omega$ decrease (see Table 3).
We can see also that for not large values of $\rho_{0}$ the quasinormal modes
of the Kaluza-Klein balck holes are longer lived (see Tables 3).

\begin{figure}[htbp]
\centering
\includegraphics[width=10cm]{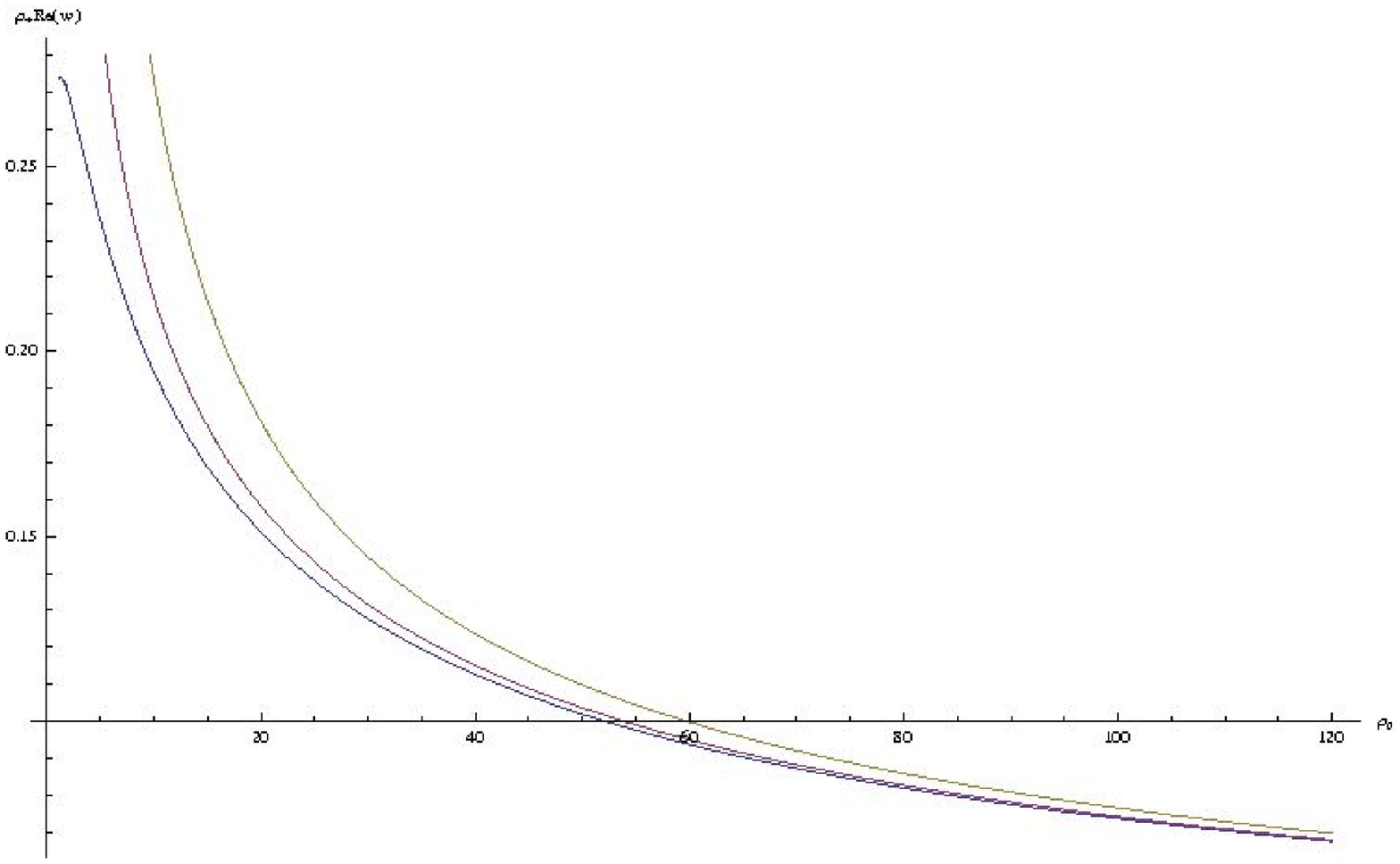}

\vspace{1cm}

\includegraphics[width=10cm]{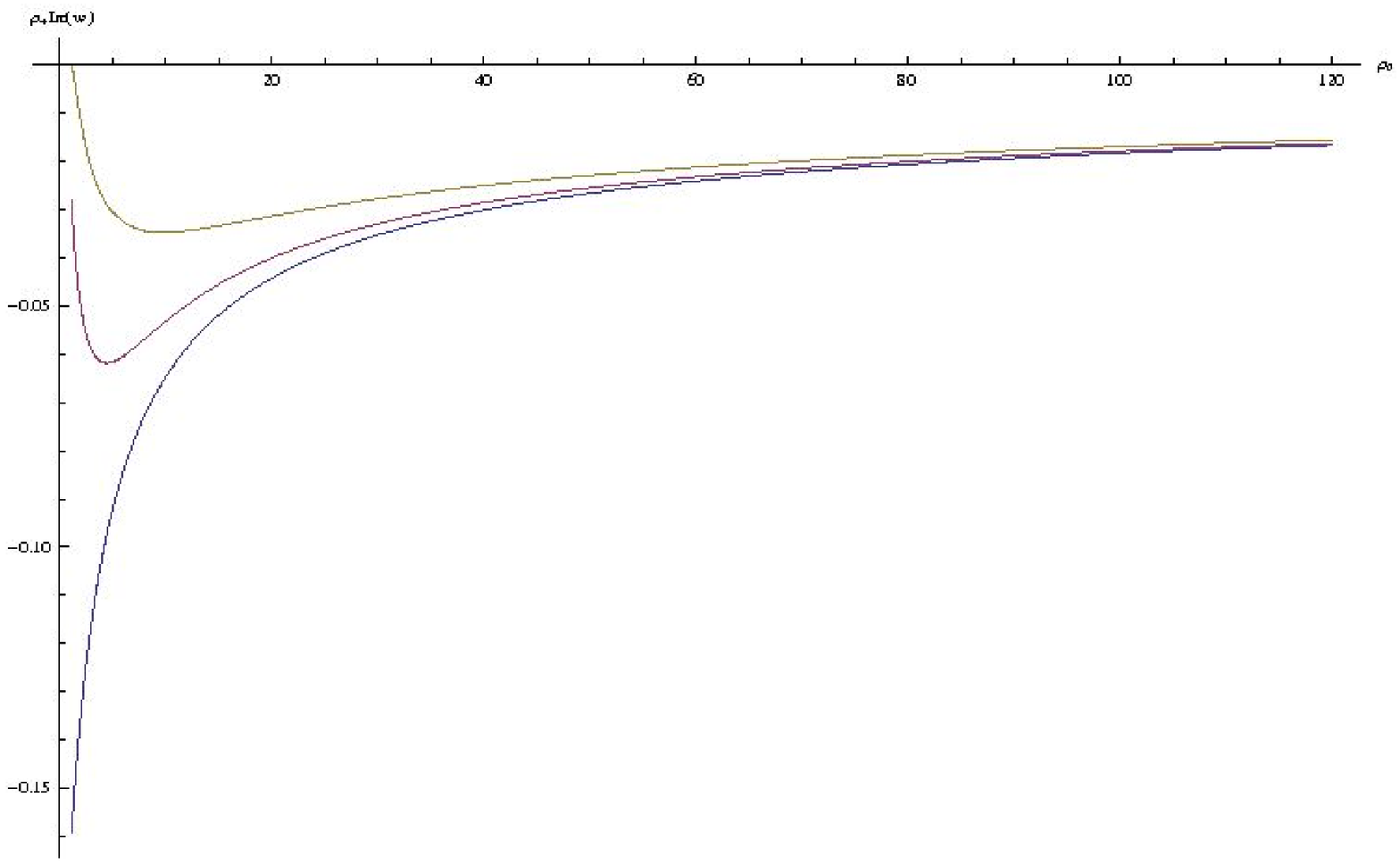}
\caption{Real and imaginary
 parts of the gravitational fundamental quasnormal modes ($n=0$) for $K=0$ (blue), $K=1$ (red), $K=2$ (yellow) modes.}
\end{figure}

\begin{table}
{\small
\begin{tabular}{|r||c|c||c|c||c|c||}
\hline
$\rho_0$&$K=0$ (t-d)&$K=0$ (WKB)&$K=1$ (t-d)&$K=1$ (WKB)&$K=2$ (t-d)&$K=2$ (WKB)\\
\hline
$ 4$&$0.2486-0.1023\imo$&$0.2467-0.1019\imo$&$0.332-0.069\imo$&$0.323-0.062\imo$&$0.506-0.026\imo$&$0.505-0.028\imo$\\
$ 6$&$0.2259-0.0840\imo$&$0.2255-0.0842\imo$&$0.282-0.068\imo$&$0.269-0.060\imo$&$0.380-0.031\imo$&$0.381-0.033\imo$\\
$ 8$&$0.2080-0.0723\imo$&$0.2083-0.0728\imo$&$0.249-0.064\imo$&$0.236-0.057\imo$&$0.313-0.034\imo$&$0.315-0.035\imo$\\
$10$&$0.1943-0.0644\imo$&$0.1944-0.0647\imo$&$0.226-0.058\imo$&$0.215-0.053\imo$&$0.272-0.035\imo$&$0.273-0.035\imo$\\
$12$&$0.1828-0.0585\imo$&$0.1828-0.0586\imo$&$0.210-0.051\imo$&$0.198-0.050\imo$&$0.240-0.034\imo$&$0.244-0.034\imo$\\
$15$&$0.1686-0.0518\imo$&$0.1688-0.0519\imo$&$0.189-0.050\imo$&$0.180-0.046\imo$&$0.212-0.033\imo$&$0.214-0.033\imo$\\
$20$&$0.1506-0.0436\imo$&$0.1511-0.0443\imo$&$0.166-0.045\imo$&$0.158-0.040\imo$&$0.181-0.027\imo$&$0.181-0.031\imo$\\
$25$&$0.1382-0.0381\imo$&$0.1380-0.0391\imo$&$0.150-0.039\imo$&$0.143-0.036\imo$&$0.160-0.024\imo$&$0.160-0.030\imo$\\
$40$&$0.1143-0.0299\imo$&$0.1127-0.0302\imo$&$0.120-0.032\imo$&$0.115-0.029\imo$&$0.122-0.021\imo$&$0.124-0.025\imo$\\
$60$&$0.0953-0.0241\imo$&$0.0938-0.0242\imo$&$0.100-0.025\imo$&$0.095-0.023\imo$&$0.098-0.018\imo$&$0.100-0.021\imo$\\
\hline
\end{tabular}
}
\caption{Quasnormal frequencies of K=1 and K=2 gravitational perturbations
of the non-rotating uncharged squased Kaluza-Klein black holes.
Frequencies are measured in units of $\rho_+$.
}
\end{table}

\begin{figure}[htbp]
\centering
\includegraphics[width=10cm]{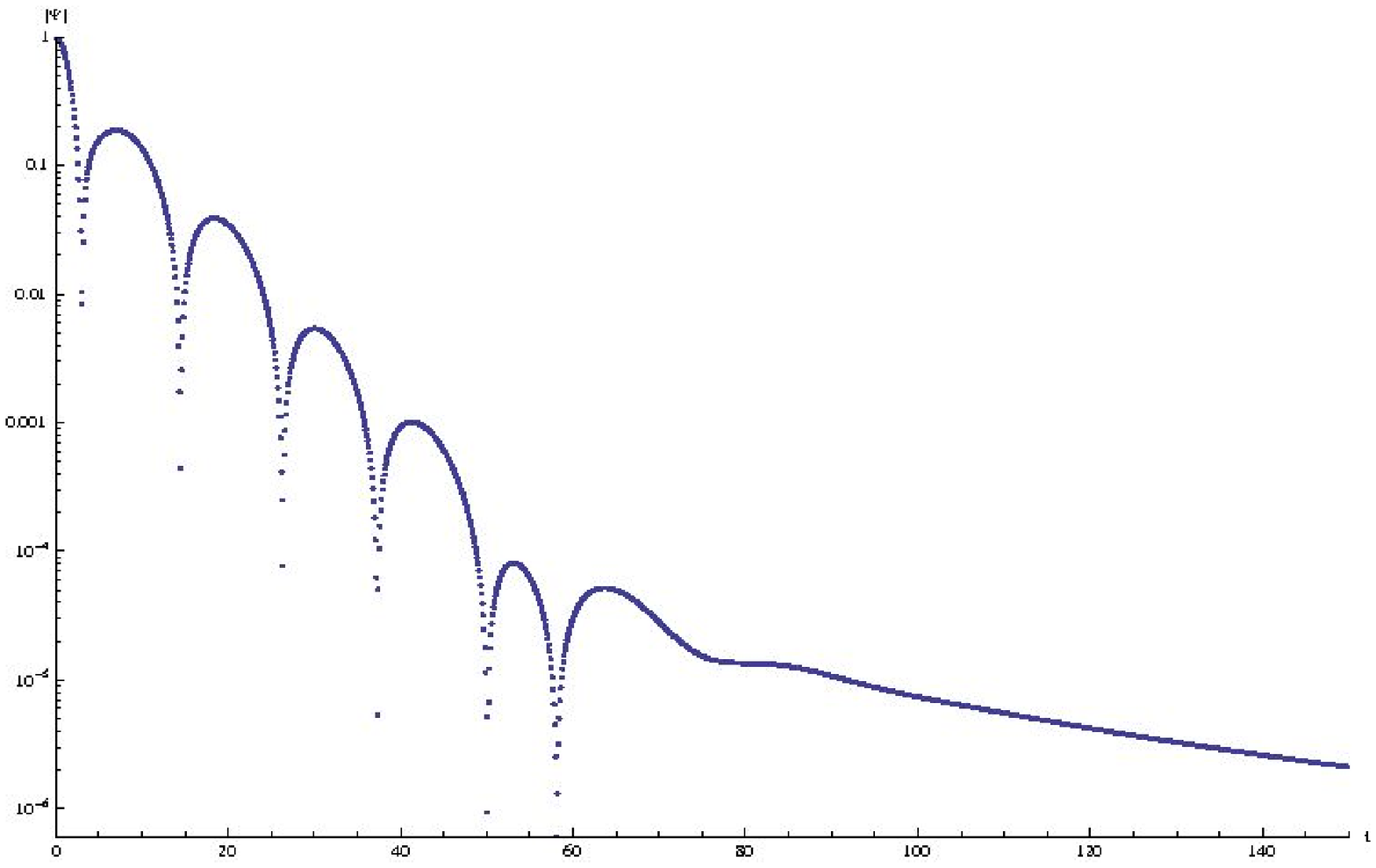}
\caption{Time domain profile for gravitational perturbations
$K=0$,
$\rho_{0} = \rho_{+} = 1$.
}
\end{figure}

\begin{figure}[htbp]
\centering
\includegraphics[width=10cm]{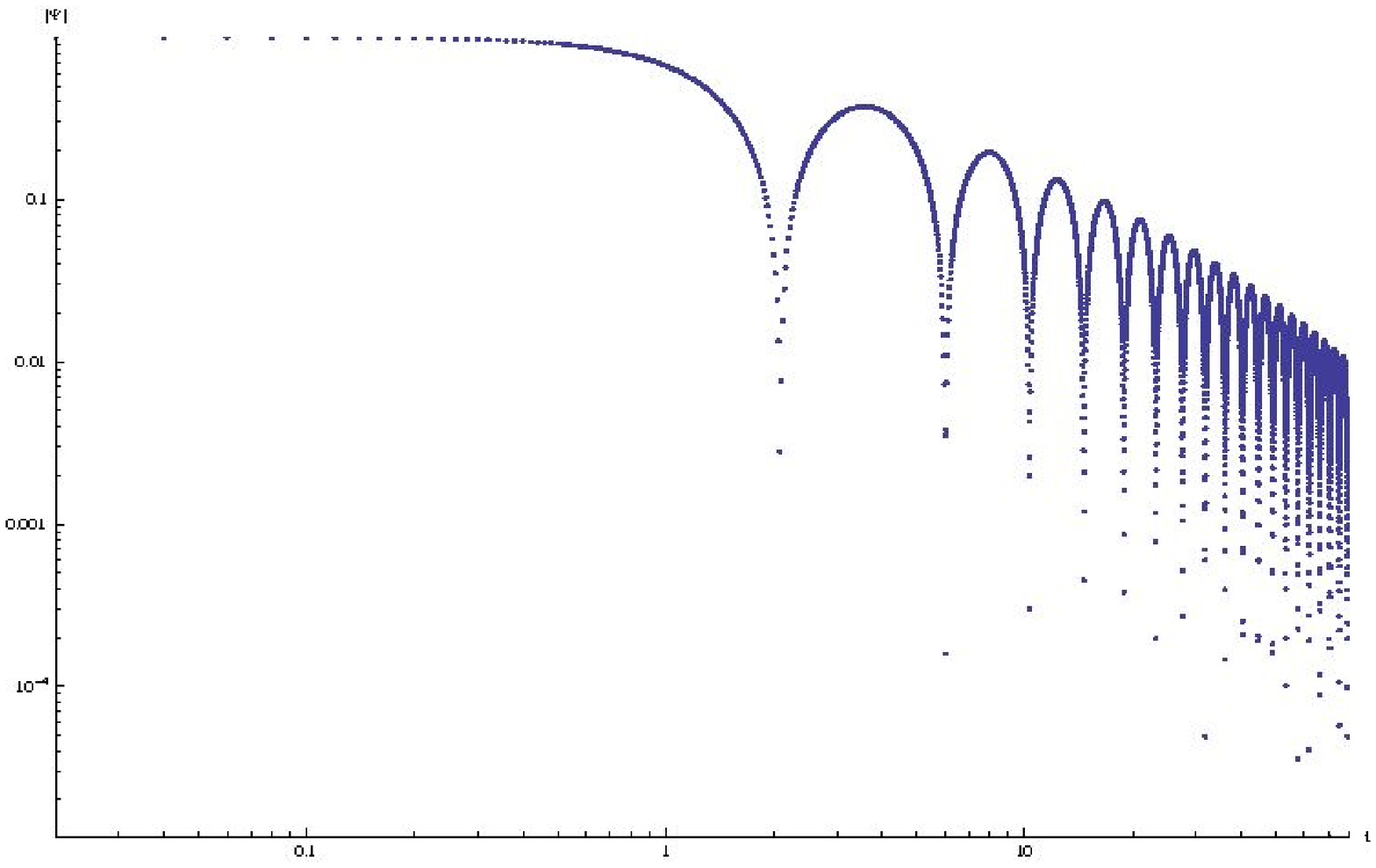}
\caption{Time domain profile for gravitational perturbations
$K=1$,
$\rho_{0} = \rho_{+} = 1$.}
\end{figure}

\begin{figure}[htbp]
\centering
\includegraphics[width=10cm]{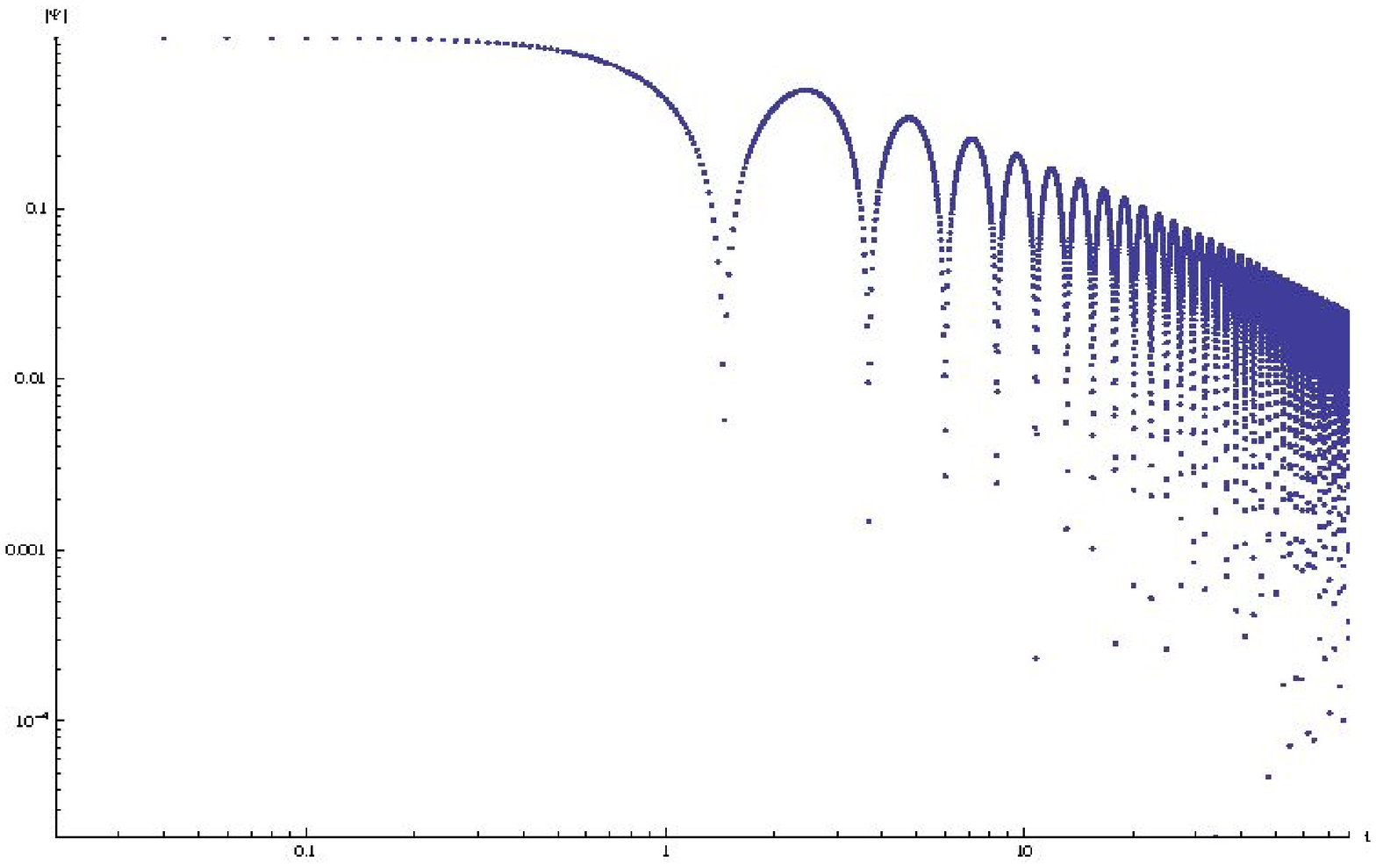}
\caption{Time domain profile for gravitational perturbations
$K=2$,
$\rho_{0} = \rho_{+} = 1$.}
\end{figure}

%
%

%

%

%


%

%

\section{Discussion}

In the present paper we considered the quasinormal spectrum of the scalar
and gravitational perturbations for squashed Kaluza-Klein black holes.
The quasinormal frequencies of the scalar field are different from those of
the Schwarzschild black holes and have smaller real oscillation frequencies
and longer lived for not very larger $\rho_{0}$.
The obtained evolution of gravitational perturbations shows damped oscillations
in time domain, so that no instability is observed
in concordance with analytically proved stability in \cite{Soda1}. Let us note that we analyzed here only ``zero mode'' perturbations, so that, strictly speaking,
the opportunity for instability is remained in higher multipole perturbations. Yet we do not expect instability at higher multipoles because, unlike $K=1$ mode,
$K=2$ mode does not have negative gap, so that higher $K$ seem simply to increase the height of the potential barrier and stabilize the system.
It is interesting that if we know the quasinormal frequency we can find the size of the extra dimension in the considered black hole model, so that quasinormal modes give
a kind of opportunity to "look into" an extra dimension at low energies. In detail, when a dominant quasinormal mode is measured, then one can look into tables and plot
numerical data in this paper and find out which is the value of $\rho_{0}$ and the radius of the event horizon that corresponds to the observed quasinormal mode. 
In this way we can determine the parameters of the black hole and the size of the extra dimension, assuming that there exists no other Kaluza-Klein black holes with similar features.

Our next step in the line of this research is to consider gravitational
perturbations of the rotating squashed Kaluza-Klein black holes.
Charged Kaluza-Klein black holes, especially maximally charged cases,
are also interesting targets for investigating quasinormal modes,
because the quasinormal frequencies of the Maxwell field and the
gravitational field coincide in four-dimensional super-symmetric
black holes\cite{OOMI}.
To check the stability of these black holes is an important problem.
As the potential is frequency dependent and cumbersome, one definitely needs
quasinormal modes analysis to test stability.

\section*{Acknowledgments}
A. Z. acknowledges support by \emph{Funda\c{c}\~ao de Amparo \`a Pesquisa do Estado de S\~ao Paulo (FAPESP)}, Brazil.
K. M. was supported in part by JSPS Grant-in-Aid for Scientific Research,
No.193715.
R. K. was supported by {\it Japan Society for Promotion of Science (JSPS)}, Japan.
H.I. and J. S. was supported by JSPS Grand-in-Aid for Scientific Research (C)
No.19540305 and No. 18540262, respectively.
J. S. also thanks the KITPC for hospitality during the period
when a part of the work on this project was carried out.



\end{document}